\documentclass{jps-cp}
\usepackage{txfonts} %Please comment out this line unless the txfonts package is availabe in your LaTeX system.
\usepackage{graphicx}        % standard LaTeX graphics tool                             % when including figure files

\title{Spin Effects In Heavy Ion Collisions at High Energies}
\author{\textsc{Zuo-tang Liang}}
\inst{Key Laboratory of Particle Physics and Particle Irradiation (MOE), Institute of Frontier and Interdisciplinary Science, Shandong University, Qingdao, Shandong 266237, China}

\email{liang@sdu.edu.cn}

\recdate{January 7, 2022}
\abst
{In non-central high energy heavy ion collisions, the colliding system possesses a huge orbital angular momentum along the normal direction of the reaction plane. 
Due to the spin orbit interaction in the system, such a huge orbital angular momentum leads to the spin polarization of quarks and anti-quarks in the quark matter system produced in the collision. 
Such an effect, known as the global polarization effect, was predicted many years ago and has been confirmed by the STAR collaboration at RHIC. 
The discovery of the global polarization effect opens a new avenue in heavy ion physics in general and in studying the properties of quark-gluon plasma in particular. 
This talk will briefly review the original ideas and calculations that lead to the prediction and summarize progresses and problems in related aspects.}
\kword{GPE (global polarization effect), HIC, QGP, spin-orbit coupling, vorticity}

\begin{document}
\maketitle

\section{Introduction}

In 2017, the global polarization effect (GPE) of $\Lambda$ and $\bar\Lambda$ hyperons 
in heavy-ion collisions (HIC) has been observed~\cite{STAR:2017ckg} by the STAR Collaboration at RHIC. 
The discovery confirms the theoretical prediction~\cite{Liang:2004ph} made more than ten years 
ago and has attracted much attention on the study of spin effects in HIC at high energies. 
This opens a new avenue to study properties of the quark-gluon plasma (QGP) and a new direction
in high energy heavy ion physics --- Spin Physics in HIC. 
Much progresses have been made and a number of reviews have been published recently. 
Among them we have in particular a special volume in the book series Lecture Notes in Physics published this year~\cite{book}. 
The field develops very fast now and it is impossible for me to cover all different aspects in this talk. 
I therefore choose to concentrate on the original ideas and calculations~\cite{Liang:2004ph,Liang:2004xn,Gao:2007bc,Gao:2020lxh} 
that lead to the predictions and a discussion of new opportunities and challenges.

Spin, as a fundamental degree of freedom of elementary particles, 
plays a very important role in modern physics and often brings us surprises.  
High energy spin physics experiments started since 1970s. 
Soon after the beginning, a series of striking spin effects have been observed 
that were in strong contradictions to the theoretical expectations at that time
and such discoveries have been pushing the studies move forward constantly.  

At the same time, high energy HIC physics has become the other active frontier in strong interaction physics 
in particular after the QGP has been discovered at RHIC.  
The study on properties of QGP in HIC is the core of high energy HIC physics currently. 

We recall that RHIC is not only the first relativistic heavy ion collider in the world 
but also the first polarized high energy proton-proton collider. 
It is therefore natural to ask whether we can do spin physics in HIC.  
Spin physics in HIC was however used to be regarded as difficult or impossible because 
the polarization of the nucleon in a heavy nucleus is very small even if the nucleus is completely polarized. 
The breakthrough came out in 2005 when it was realized that~\cite{Liang:2004ph} there is however 
a great advantage to study spin and/or angular momentum effects in HIC,  
i.e., the reaction plane in a HIC can be determined experimentally %by measuring flows and/or spectator nucleons 
and there exist a huge orbital angular momentum (OAM) for the participating system in a non-central HIC with respect to the reaction plane! 
It provides a unique place in high energy reactions to study the mutual exchange of the OAM and the spin polarization.  
The discovery of GPE leads to an active field of Spin Physics in HIC~\cite{Liang:2019clf}. 
The aim of this talk is to briefly introduce the original ideas, the main results and new challenges.

\section{{Orbital angular momenta (OAM) of QGP in HIC}}
\label{secoam}

\subsection{The reaction plane in HIC}
\label{secreactionplane}

\begin{figure}[htbp]
\begin{center}
%\resizebox{2.5in}{2.1in}{\includegraphics{geo2.eps}}
\includegraphics[width=5.0cm]{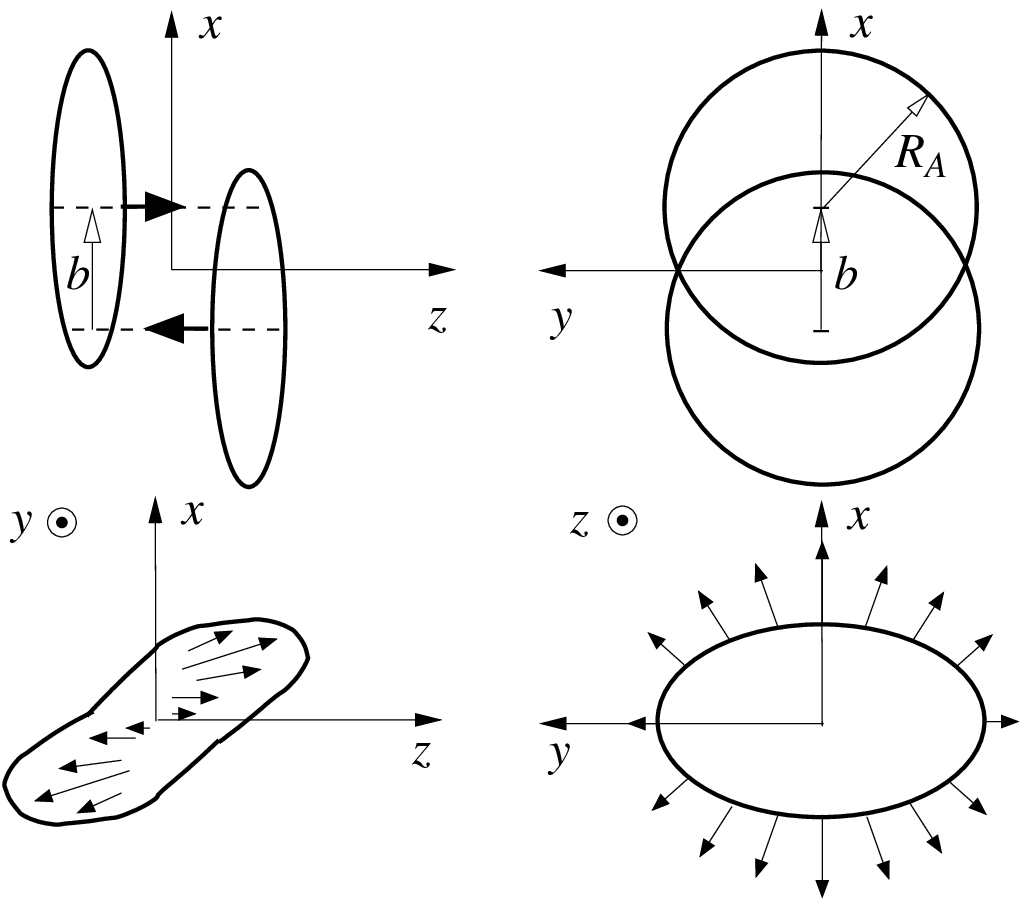} \hspace{0.7cm}
\includegraphics[width=6.7cm]{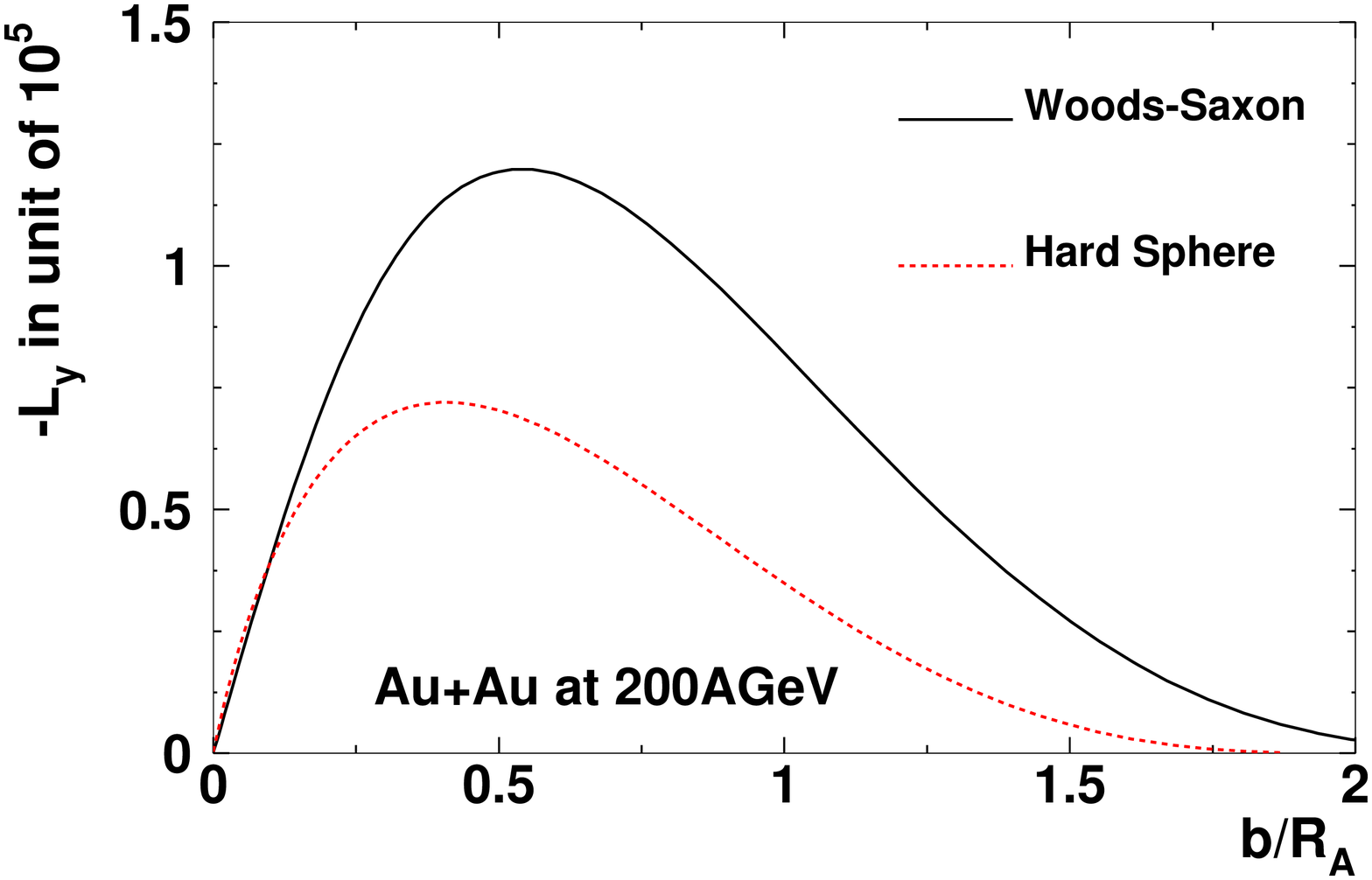}
\end{center}
\caption{Illustration of the geometry (left) and the global OAM (right) of the colliding system 
for the non-central HIC with impact parameter $\vec{b}$. }
%This figure is taken from~\cite{Liang:2004ph}. }
\label{figgeo}
\end{figure}

We consider two colliding nuclei with the projectile of beam momentum per nucleon $\vec p_{in}$.
For a non-central collision, there is a transverse separation between the centers of the two colliding nuclei. 
The impact  parameter $\vec b$ is defined as the transverse vector pointing from the target to the projectile. 
The reaction plane is defined by $\vec b$ and $\vec p_{in}$ as specified in the upper left panel in Fig.~\ref{figgeo} 
where the normal of the plane is $\vec{n} = {\vec p}_{in}\times{\vec{b}}/|{\vec p}_{in}\times{\vec{b}}|$. 
The overlap parts, hereafter referred as the colliding system, 
interact with each other and form the system in the middle while the other parts are just spectators 
that move apart in the original directions. 

Usually in a high energy reaction such as a $hh$, or $lh$ or $e^+e^-$ annihilation, 
the size of the reaction region is typically less than 1fm. 
The reaction plane in such collisions can be defined theoretically but can not be determined experimentally. 
However, in a HIC, the reaction region is usually much larger and colliding parts  give rise to a quark matter system 
with very high temperature and high density and expand violently while the spectators just leave 
the region in the original directions. 
Since the colliding system is not isotropic, the pressures in different directions are also different  
thus lead to a system that expands non-isotropically.  
In the transverse directions they behave like an ellipse as illustrated in Fig.~\ref{figgeo}. 
Such an anisotropy is described by the elliptic flow $v_2$ and the directed flow $v_1$~\cite{Ollitrault:1992bk} 
%defined as the coefficient of $\cos2\phi$ and $\cos\phi$ respectively,
that can be measured experimentally. 
Clearly, by measuring $v_2$, one can determine the reaction plane 
and further determine the direction of the plane by measuring the directed flow $v_1$. 

In experiments, the reaction plane in a HIC can be determined not only by measuring $v_2$ and $v_1$ 
but also by measuring the sidewards deflection of the forward- and backward-going 
fragments and particles in the beam–beam counter detectors~\cite{STAR:2017ckg}.  
This is quite unique in different high energy reactions. 

\subsection{The global orbital angular momentum (OAM)}

Just as illustrated in Fig.~\ref{figgeo}, in a non-central HIC, there is a transverse separation between the overlapping parts 
of the two colliding nuclei in $\vec b$ direction.
Hence the colliding system carries a finite global OAM $L_y$ along the direction orthogonal to the reaction plane.
The magnitude of $L_y$ has been calculated~\cite{Liang:2004ph,Gao:2007bc} and the results obtained are shown in Fig.~\ref{figgeo}.
We see that the global $L_y$ is huge and is of the order of $10^5$ at most impact parameters.

\subsection{The transverse gradient of the momentum distribution and the local OAM}

The global OAM could be distributed across the overlapped
region of nuclear scattering and manifest itself in the shear of
the longitudinal flow leading to a finite value of local vorticity. 
Under such longitudinal fluid shear, a pair of scattering partons on average carry a finite 
relative OAM that is referred to as the local OAM in the opposite direction of $\vec n$.  
How the global OAM is distributed to the longitudinal flow shear and 
the magnitude of the local relative OAM depends on the parton production mechanism in HIC. 
Numeric results were obtained~\cite{Liang:2004ph,Gao:2007bc}  using the Landau fireball and Bjorken scaling models. 

\begin{figure}[ht]
\begin{center}
\includegraphics[width=5.7cm]{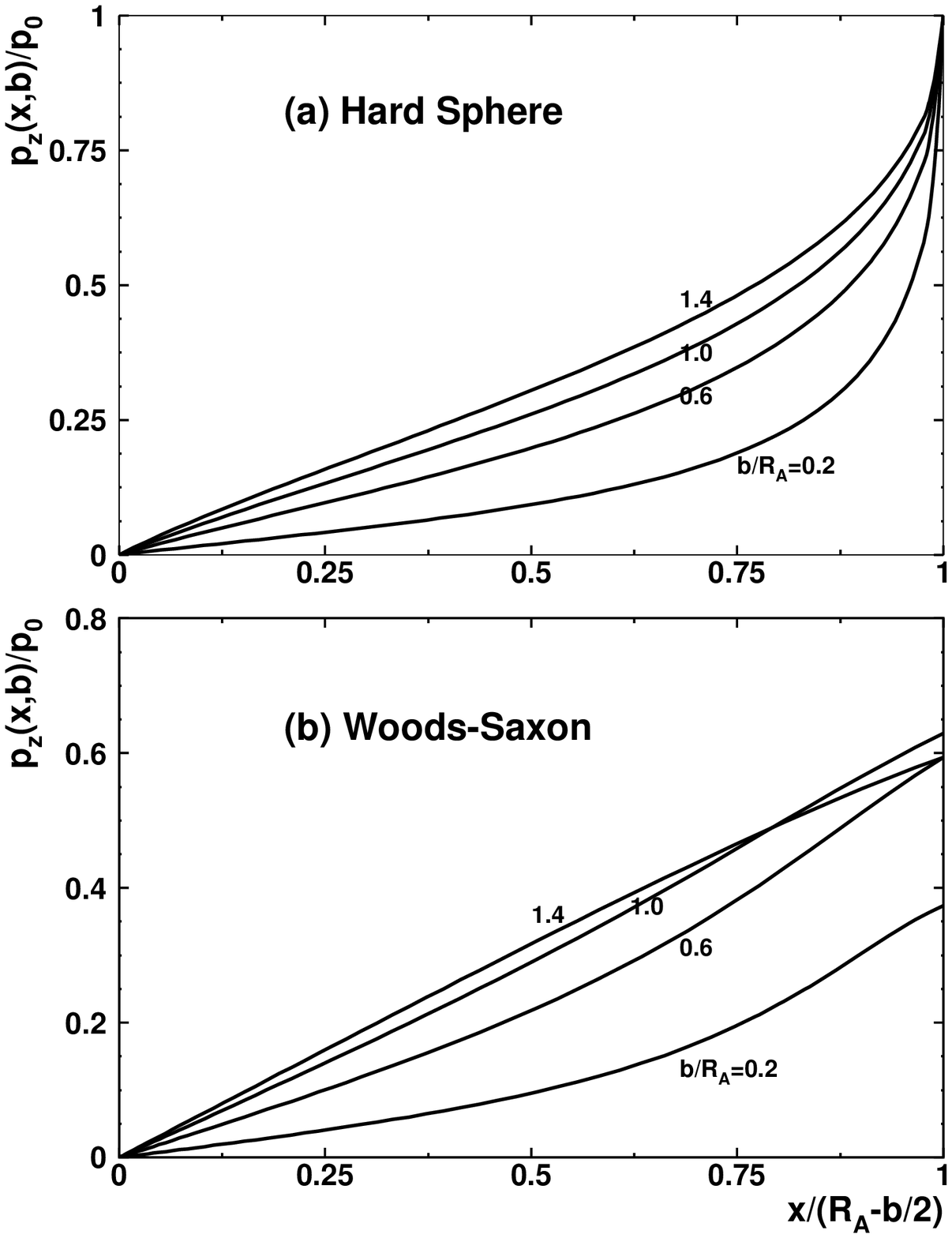} \hspace{-1cm}
\includegraphics[width=5.7cm]{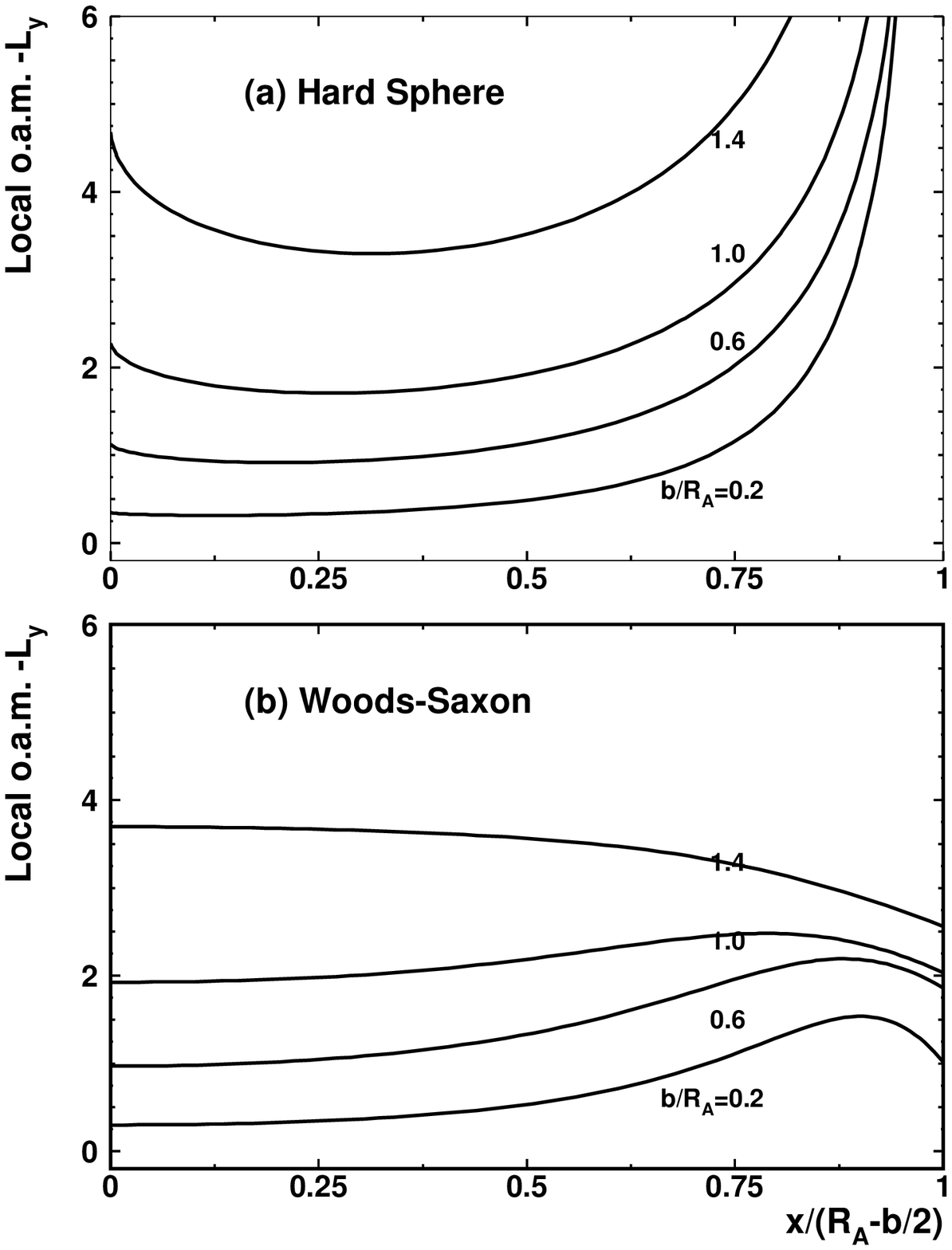}
\end{center}
\caption{The average longitudinal momentum $p_z(x,b,\sqrt{s})$ in unit of $p_0=\sqrt{s}/2c(s)$ 
and the average OAM $l_y$ of two neighboring partons separated by $\Delta x=1$ fm
as functions of $x/(R_A-b/2)$ for different values of $b/R_A$.  
$c(s)$ is the average number of partons produced per participating nucleon. }
\label{figpxb}
\end{figure}

In the Landau fireball model, we assume that the produced partons thermalize quickly 
and have a common longitudinal flow velocity at a given transverse position in the overlapped region. 
The average longitudinal momentum $p_z(x,b,\sqrt{s})$ 
and the average local OAM $l_y$ for two neighboring partons
have been calculated~\cite{Liang:2004ph,Gao:2007bc} and the results are given in Fig.~\ref{figpxb}.
In $Au+Au$ collisions at $\sqrt{s}=200$GeV, $c(s)\simeq 45$ and $dp_0/dx \simeq 0.34$ GeV/fm,  
we obtain $l_0\simeq -1.7$ for $\Delta x=1$ fm.
Hence, $l_y$ is in general of the order of 1 and is comparable or larger than the spin of a quark.

In the Bjorken scaling scenario,  
the longitudinal flow velocity is identical to the spatial velocity $\eta=\log[(t+z)/(t-z)]$. 
With such correlation, the local interaction and thermalization require that 
a parton only interacts with others in the same region of longitudinal momentum or rapidity $Y$. 
The width of such region is determined by the half-width of the thermal distribution. 
In this case, we should consider the average $\langle Y_l\rangle$ of rapidity of partons in the interval. 
The relevant measure of the local relative OAM between two interacting partons is, 
therefore, the difference in $\langle Y_l\rangle$ for partons at transverse distance of  the order of the average interaction range.

\begin{figure}[ht]
\begin{center}
\includegraphics[width=7.7cm]{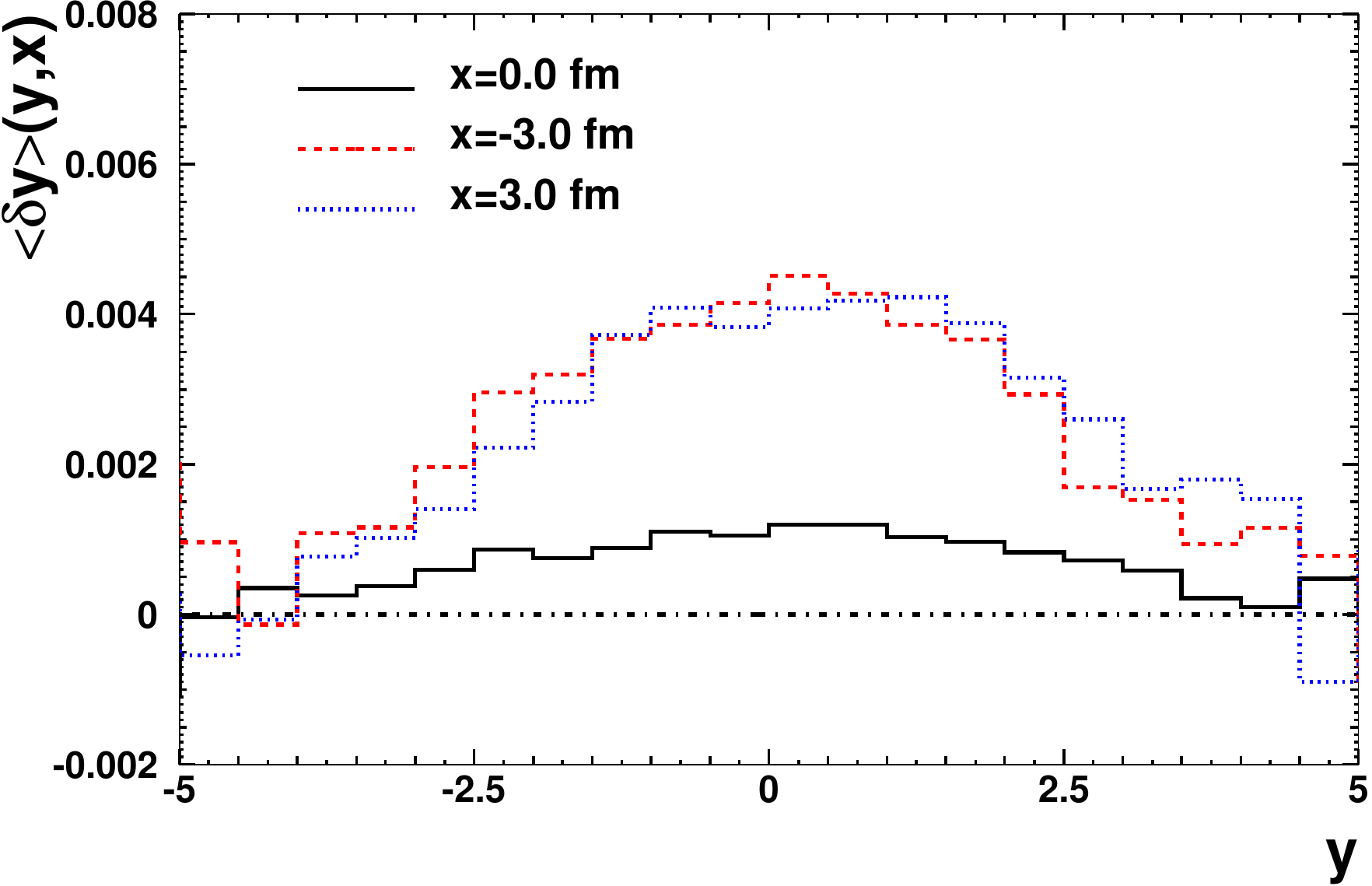}
\end{center}
\caption{The average rapidity shear $\partial \langle Y_l\rangle/\partial x$ within
a window $\Delta_Y=1$ as a function of $Y$ at
different $x$ from HIJING in non-central $Au+Au$ collisions at $\sqrt{s}=200$ GeV. 
This figure is taken from~\cite{Gao:2007bc} where the $y$-axis label is different from the notation in this talk.}
\label{figdndydxHIJING}
\end{figure}

For numerical calculations, 
one needs a dynamical model to estimate the local rapidity distribution $d^2N/dxdY$ of produced partons. 
For this purpose, the HIJING Monte-Carlo model~\cite{Wang:1991ht,Wang:1996yf} 
and the BGK model~\cite{Brodsky:1977de} were used~\cite{Gao:2007bc,Liang2019}. 
Shown in Fig.~\ref{figdndydxHIJING} is the average rapidity shear $\partial\langle Y_l\rangle/\partial x$ 
as a function of $Y$ at different $x$.
We see that the average rapidity shear has a positive and finite value in the central rapidity region. 
Such qualitative feature is the same as that obtained in the Landau fireball model.  
Quantitatively,  the corresponding local relative longitudinal momentum shear 
$\partial \langle p_z\rangle/{\partial x}\sim p_T\cosh Y ~ \partial\langle Y_l\rangle/\partial x$.  
With $\langle p_T\rangle \approx 2T\sim 0.8$ GeV, we have $\partial \langle p_z\rangle/\partial x\sim 0.003$ GeV/fm 
in the central rapidity region, which is smaller than that from a Landau fireball model estimate.

\section{Theoretical predictions on the GPE of QGP in HIC}
\label{secgp}

It has been shown~\cite{Liang:2004ph} that due to spin-orbit interactions in a strongly interacting system such as QGP, 
the relative OAM can be transferred to the polarization of the constituents in the system 
such as the quarks and anti-quarks. 
This leads to the prediction that~\cite{Liang:2004ph} the QGP produced in a high energy HIC is globally polarized 
in the direction opposite to the normal of the reaction plane and is known as the global polarization effect (GPE). 
We briefly summarize the calculations and results 
in the first papers~\cite{Liang:2004ph,Gao:2007bc,Liang:2004xn} that lead to the prediction in the following. 
A comprehensive review can be found in~\cite{Gao:2020lxh}.

\subsection{Global quark polarization in QGP in HIC}
\label{secqpol}

We have seen that in a non-central $AA$ collision, 
there is a huge global OAM for the colliding system. 
Such a global OAM leads to the longitudinal fluid shear in the produced system of partons. 
A pair of interacting partons will have a finite value of relative
OAM along the direction opposite to the normal of the reaction plane. 
It is thus natural to ask whether the OAM or momentum shear lead to the polarization of partons in the system. 

There is however no field theoretical calculation that can be applied directly to answer this question 
because usually the calculations are carried out in the momentum space where the momentum shear with respect to $x$ coordinate 
can not be taken into account. 
To achieve this, Ref.~\cite{Liang:2004ph} took the approach by considering parton scattering with impact parameter in the preferred direction. 
The calculations have been carried out in Refs.~\cite{Liang:2004ph,Gao:2007bc}. 
We present the main outline and conclusions in the following. 

\subsubsection{Quark scattering at fixed impact parameter}

We consider the scattering $q_1(p_1)+q_2(p_2)\to q_1(p_3)+q_2(p_4)$ of two quarks with different flavors.
The scattering matrix element in momentum space is given by,
\begin{equation}
S_{fi}=\langle f|\hat S|i\rangle={\cal M}_{fi}(q) (2\pi)^4\delta^4(p_1+p_2-p_3-p_4),
\label{eqsfi}
\end{equation}
where ${\cal{M}}_{fi}(q)$ is the scattering amplitude with momentum transfer $q$ in the momentum space. 
It has been shown that the differential cross section can be written as,
%\begin{eqnarray}
%&&d\sigma=\frac{c_{qq}}{F}
%\frac{|S_{fi}(q)|^2}{TV}\frac{d^3{{p}}_3}{(2\pi)^{3}2E_3} \frac{d^3{{p}}_4}{(2\pi)^{3}2E_4},~~~~~~
%\label{eqqqcs}
%\end{eqnarray}
\begin{equation}
d\sigma=\frac{c_{qq}}{F}\int d^2x_T\int\frac{d^2q_\perp}{(2\pi)^2}\frac{d^2k_\perp}{(2\pi)^2} e^{-i(\vec q_T-\vec k_T)\cdot\vec x_T}
\frac{{\mathcal{M}}_{fi}(q)}{\Lambda(q)}\frac{{\mathcal{M}}_{fi}^{*}(k)}{\Lambda(k)} ,~~~
\label{eqcsxT}
\end{equation}
where $c_{qq}$ is the color factor, $F$ is the flux factor and 
${\Lambda(q)} =\sqrt{(E_1+E_2)p_{3z}}$ is a kinematic factor. 
Hence, the differential cross section at given impact parameter is given by 
\begin{equation}
\frac{d^2\sigma_{\lambda_3}}{d^2x_T}
=\frac{c_{qq}}{16F}\sum_{\lambda_1,\lambda_2,\lambda_4} \int\frac{d^2q_T}{(2\pi)^2}\frac{d^2k_T}{(2\pi)^{2}}
e^{i({\vec{k}}_{T}-{\vec{q}}_{T})\cdot{\vec{x}}_{T}}
\frac{\mathcal{M}(q)}{\Lambda(q)} \frac{{\mathcal{M}}^{*}(k)}{{\Lambda}(k)} ,
\label{eqdcslambdadxT}
\end{equation}
where $\lambda_i=+$ or $-$ is the spin of $q_i$.  
Averaging over the distribution $f_{qq}(\vec x_T,b,Y,\sqrt{s})$ of $\vec x_T$, we obtain that the polarization of $q_1$ after the scattering is given by, 
\begin{eqnarray}
&&P_q=\langle\Delta\sigma\rangle/\langle\sigma\rangle, \label{eqPqDcsovercs} \\
&&\langle\Delta\sigma\rangle=\int d^2x_T f_{qq}(\vec x_T,b,Y,\sqrt{s}) \frac{d^2\Delta\sigma}{d^2x_T},  ~~~~~~~~ %\label{eqDcsint} \\&&
\langle\sigma\rangle=\int d^2x_T f_{qq}(\vec x_T,b,Y,\sqrt{s}) \frac{d^2\sigma}{d^2x_T},  \label{eqcsint}
\end{eqnarray}
where the unpolarized ${d^2\sigma}/{d^2x_T}$ and spin-dependent part ${d^2\Delta\sigma}/{d^2x_T}$ of the cross section are the sum and the difference of  
${d^2\sigma_{\lambda_3}}/{d^2x_T}$ for $\lambda_3=+$ and $-$. 

The distribution $f_{qq}(\vec x_T,b,Y,\sqrt{s})$ 
depends on the collective longitudinal momentum distribution shown in Sec.~\ref{secoam}.
Clearly, it should be determined by the dynamics of QGP and HIC. 
To see the qualitative features of physical consequences, 
Refs.~\cite{Liang:2004ph,Gao:2007bc} took a simplified form $f_{qq}(\vec x_T,b,Y,\sqrt{s})\propto \theta(x) $, i.e.,  
a uniform distribution of $\vec x_T$ in the upper half $xy$-plane.

\subsubsection{Quark scattering by a static potential}
\label{csSPM}

To see the characteristics of the physical consequences clearly, Ref.~\cite{Liang:2004ph} 
started with the quark scattering by a static potential. 
In this case, we have,
\begin{equation}
{\cal M}_{fi}(q)=\bar{u}_{\lambda}(p+q)~ A\hspace{-5pt}\slash (q) ~ u(p), ~
\label{eqmSPM}
\end{equation}
where $A(q)=(A_0(q),\vec 0)$ and $A_0(q)=g/(q^2+\mu_D^2)$ is the screened static potential
with Debye screen mass $\mu_D$. %~\cite{gw93}.
For small angle scattering, $q_T\sim k_T\sim \mu_D \ll E$, the energy of the quark, we obtain~\cite{Liang:2004ph},
\begin{eqnarray}
&&\frac{d^2\sigma}{d^2x_T}=\alpha_s^2c_T K_0^2(\mu_{D}x_T),  ~~~~~~~ %\label{eqcsresspm} \\ 
\frac{d^2\Delta\sigma}{d^2 x_T}=\alpha_s^2c_T
\left[(\vec{p}\times\vec{n})\cdot\hat{\vec{x}}_T/\vec p^2\right] \mu_D K_0(\mu_{D}x_T)K_1(\mu_{D}x_T). \label{eqDcsresspm}
\end{eqnarray}
where $J_0$ and $K_0$ are the Bessel and modified Bessel functions respectively; $x_T=|\vec x_T|$ and $\hat{\vec x}_T=\vec x_T/x_T$.
Averaging over $\vec{x}_T$,  one obtains the global quark polarization,
\begin{equation}
P_q = - \pi\mu_D|\vec p|/2E(E+m_q).
\label{eqPqSPM}
\end{equation}
%via a single scattering for given $E$. 
%
If one takes the non-relativistic limit, $E\sim m_q \gg |\vec p|$ and $\mu_D$, one obtains, 
$P_q\approx - \pi \mu_D |\vec p| /4m_q^2$. 
%\label{eqPqSPMnr}
%\end{equation}
It can be verified that~\cite{Liang:2004ph} this is just the result due to spin-orbit interaction.

We also note that the cross sections can be written in a general form as,
\begin{eqnarray}
\label{eqcsgform}
&&\frac{d^2\sigma}{d^2 x_{T}}=F(x_T,E),  ~~~~~~~
\frac{d^2\Delta\sigma}{d^2x_{T}}
=\vec{n}\cdot(\hat{\vec{x}}_T\times{\vec{p}}\ )~\Delta F(x_T,E),
\end{eqnarray}
where $F(x_T,E)$ and $\Delta F(x_T,E)$ are scalar functions of $x_T$ and $E$.

\subsubsection{Quark-quark scattering in a thermal medium}
\label{csqqhtl}

The quark-quark scattering amplitude in a thermal medium can be calculated 
by using the Hard Thermal Loop resumed gluon propagator. %~\cite{WELD82,hw96}. 
The calculations are much more involved and are given in~\cite{Gao:2007bc}. 
The results obtained under the small angle approximation are~\cite{Gao:2007bc}
\begin{eqnarray}
&&\frac{d^2\sigma}{d^2 x_T}= \frac{c_{qq}}{2}\alpha_s^2
\left[K_0(\mu_{m}x_T)+K_0(\mu_{D}x_T)\right]^{2}, \label{eqcsreshtl} \\
&&\frac{d^2\Delta\sigma}{d^2 x_T} =
\frac{c_{qq}\alpha_s^2}{2\vec p^2}
\left[(\vec{p}\times\vec{n})\cdot\hat{\vec x}_T\right] \left[K_0(\mu_{m}x_T)+K_0(\mu_{D}x_T)\right] \bigl[\mu_{m}K_1(\mu_{m}x_T)+\mu_{D}K_1(\mu_{D}x_T)\bigr]. \label{eqDcsreshtl}
\end{eqnarray}
They are similar to those given by Eq.~(\ref{eqDcsresspm}) for the screened static potential model.
The only differences are additional contributions from magnetic gluons that are absent in the static potential model.

\begin{figure}[htbp]
\begin{center}
\includegraphics[width=7.7cm]{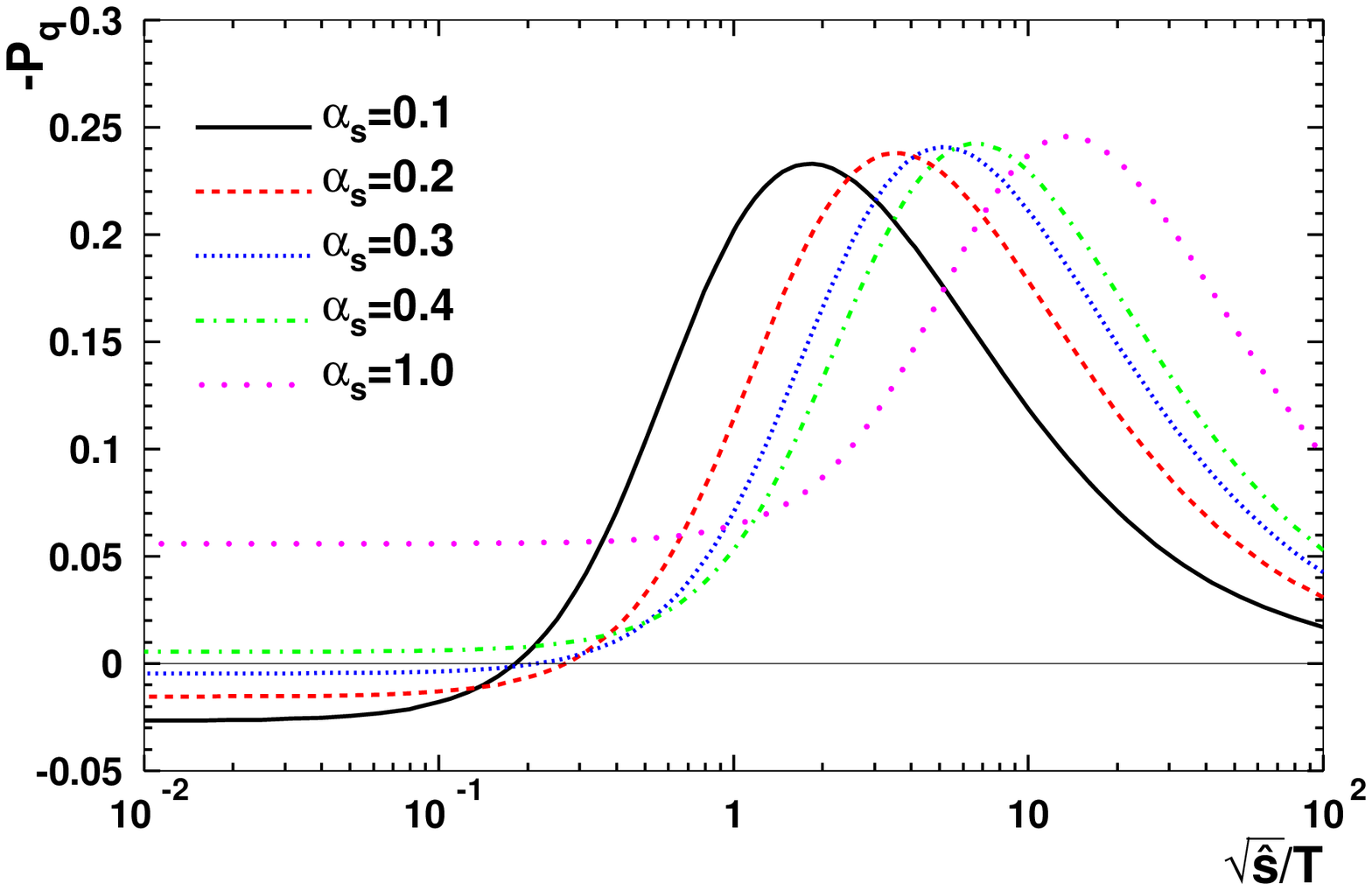}
\end{center}
%\centerline{\psfig{figure=geo2.eps,width=3.0in,height=2.4in}}
\caption{Quark polarization -$P_q$ as a function of
$\sqrt{\hat{s}}/T$ for different $\alpha_s$'s obtained in quark-quark scattering with a hard thermal loop resumed gluon propagator. }
%This figure is taken from~\cite{Gao:2007bc}. }
\label{figqpolhtl}
\end{figure}

Going beyond the small angle approximation, we obtain~\cite{Gao:2007bc} expressions much more complicated but take the same form 
as that given by Eq.~(\ref{eqcsgform}). 
The numeric results of quark polarization $P_q$ as function of c.m. energy $\sqrt{\hat{s}}/T$ of the $qq$-system is shown in Fig.~\ref{figqpolhtl}. \\[-3cm]

\subsubsection{Conclusions and discussions on the global quark polarization}

Although approximations and/or models have to be used in the calculations presented above, 
the physical picture and the consequence are very clear. 
It is confident that after the scattering of two constituents in QGP, 
the OAM will be transferred partly to the polarization of quarks and anti-quarks due to spin-orbit coupling in QCD. 
Such a polarization is different from those that we discussed before in high energy physics
such as the longitudinal or transverse polarization where directions are defined w.r.t. the momentum of the produced hadron 
and are in general different for different hadrons in the same event.  
In contrast, the polarization discussed here refers to the normal of the reaction plane  
that is fixed for one event and is independent of any particular hadron in the final state. 
Hence, in Ref.~\cite{Liang:2004ph}, this polarization was given a new name --- the global polarization, 
and the QGP was referred to the globally polarized QGP in non-central HIC. 

The following three points should be addressed in this connection. 

(i) %(Multiple scattering and equilibrium) 
The results presented above are obtained for a single quark-quark scattering. 
In a realistic HIC where QGP is created, such quark-quark scatterings may take place for a few times before the hadronization. 

(ii) %(Pol rate and local OAM) 
The numerical results on quark polarization presented above are obtained using the simple form $f_{qq}(\vec x_T,Y,b,\sqrt{s}~)\propto \theta (x)$. 
They provide a practical guidance for the magnitude of the quark polarization but can not give 
us the relationship between the polarization and the local OAM. 
In practice, to describe the evolution of the global quark polarization, 
one needs a dynamical model of QGP evolution or effectively a dynamical model for $f_{qq}(\vec x_T,Y,b,\sqrt{s}~)$. 

(iii) %(Vorticity) 
If we consider QGP as a fluid, the momentum shear distribution discussed in Sec.~\ref{secoam} 
implies a non-vanishing vorticity. % $\vec\omega=(1/2) \nabla\times\vec v$. 
The spin-orbit coupling can be replaced by spin-vortical coupling and 
the GPE provides a good opportunity to study spin-vortical effects in strongly interacting system.  
This was first discussed in~\cite{Betz:2007kg} and attracted much attention. 
Much progresses have been made in particular that the equilibrium has been introduced~\cite{Becattini:2007sr,Becattini:2013fla,Becattini:2016gvu} 
to derive the relationship between vorticity and spin polarization
and that Monte-Carlo event generators have been used to simulate the vorticity of QGP~\cite{Deng:2016gyh,Pang:2016igs}. 
For a comprehensive review, see~\cite{book} and references cited therein.

\subsection{Global hadron polarization in HIC}
\label{sechpol}

The most obvious and measurable effect from 
the global quark polarization in QGP in non-central HIC is the global polarization of hadrons produced after the hadronization.  
The global hyperon polarization and vector meson spin alignment are given in~\cite{Liang:2004ph} and~\cite{Liang:2004xn}.
The results depend clearly on the hadronization mechanism. 
We briefly summarize here. 

\subsubsection{Global hyperon polarization}

(i) {\it In the quark combination mechanism}:  
We assume that there is no correlation between polarizations of different quarks and obtain 
the polarization of the hyperon $H$ in table~\ref{tabHpol}. 
It is obvious that if $P_u=P_d=P_s\equiv P_q$, we have $P_H=P_q$ for all hyperons. 

\begin{table}
\caption{Polarizations of hyperons produced in the quark combination or fragmentation mechanism. 
 The results for fragmentation are for leading hadrons only where $n_s$ and $f_s$ 
 are strange quark abundances relative to up or down quarks in QGP and quark fragmentation, respectively.
%These results are taken from~\cite{Liang:2004ph}.
} \label{tabHpol}
\begin{tabular}{ccccccc}\hline
hyperon & $\Lambda$ & $\Sigma^+$ & $\Sigma^0$ &$\Sigma^-$ &$\Xi^0$ &$\Xi^-$  \rule[-0.20cm]{0mm}{0.6cm} \\ \hline 
combination &~~~$P_s$~~~ & $\frac{4P_u-P_s}{3}$& $\frac{2(P_u+P_d)-P_s}{3}$ 
&$\frac{4P_d-P_s}{3}$&$\frac{4P_s-P_u}{3}$ & $\frac{4P_s-P_d}{3}$ \rule[-0.25cm]{0mm}{0.7cm} \\ \hline 
fragmentation &~~~$\frac{n_sP_s}{n_s+2f_s}$ &
~~$\frac{4f_sP_u-n_sP_s}{3(2f_s+n_s)}$ & ~~$\frac{2f_s(P_u+P_d)-n_sP_s}{3(2f_s+n_s)}$ & ~~$\frac{4f_sP_d-n_sP_s}{3(2f_s+n_s)}$ &
~~$\frac{4n_sP_s-f_sP_u}{3(2n_s+f_s)}$& ~~$\frac{4n_sP_s-f_sP_d}{3(2n_s+f_s)}$ 
\rule[-0.25cm]{0mm}{0.7cm} \\ \hline 
\end{tabular}
\end{table}

(ii) {\it In the quark fragmentation mechanism}:  
In this case, the hyperon polarization is described by polarized quark fragmentation functions.
In \cite{Liang:2004ph}, an estimation was made for the polarization of leading hyperons and the results are also given in table~\ref{tabHpol}.
We see if $n_s=f_s$ the result from fragmentation is just $1/3$ of the corresponding result from combination.

\subsubsection{Global spin alignment of vector mesons}

Vector meson spin alignment is given by the $00$-component $\rho^V_{00}$ of the spin density matrix. 

(i) {\it In the quark combination mechanism}:  
If we do not consider the correlation between polarizations of quarks and anti-quarks, 
we obtain $\rho^V_{00}$ as~\cite{Liang:2004xn},
\begin{equation}
\rho^V_{00}%=\frac{ (1+P_{q_1})(1-P_{\bar q_2})+  (1-P_{q_1})(1+P_{\bar q_2})}{3+P_{q_1}P_{\bar q_2}}
=({1-P_{q_1}P_{\bar q_2}})/({3+P_{q_1}P_{\bar q_2}}) . \label{eqrhoV00}
\end{equation}
We see that $\rho^V_{00}$ is a quadratic effect of $P_q$ and should be less than $1/3$.   
 
(ii) {\it In the quark fragmentation mechanism}:  
The spin alignment $\rho_{00}$ is described by the $S_{LL}$-dependent fragmentation function $D_{1LL}(z)$.  
A parameterization is obtained in~\cite{Chen:2016iey} by fitting to data available~\cite{Ackerstaff:1997kj,Abreu:1997wd} in $e^+e^-$ annihilations. 
The result shows that,  in contrast to quark combination, $\rho_{00}$ obtained in fragmentation is larger than $1/3$. 
This implies that the spin of $\bar q$ produced in the fragmentation $q\to h+X$ has larger probability to be in the opposite direction as $q$. 
For the leading meson, a parameterization of $P_{\bar q}=-\beta P_q$ %(where $\beta\sim 0.5$) 
for the $\bar q$ produced in the fragmentation and combine with the initial $q$ to form the leading vector meson was proposed~\cite{Xu:2001hz} to fit the data~\cite{Ackerstaff:1997kj,Abreu:1997wd}.
Ref.~\cite{Liang:2004xn} also made an estimation for such leading vector mesons in fragmentation based on this empirical relation and obtained that,  
\begin{equation}
\rho^V_{00}=(1+ \beta P_q^2)/(3-\beta P_q^2) . \label{eqrhoV00phe}
\end{equation}
We see that the spin alignment $\rho^V_{00}$ obtained this way is indeed larger than $1/3$.

\section{Comparison with experiments}

\subsection{Experimental results for global hyperon polarizations}

The novel predictions~\cite{Liang:2004ph,Gao:2007bc,Liang:2004xn} on GPE have attracted immediate attentions also experimentally. 
Shortly after the publication of theoretical predictions, 
the STAR collaboration had started measurements both on $\Lambda$ polarization and vector meson spin alignments at $\sqrt{s}=200$ GeV.  
However, results of early measurements~\cite{Abelev:2007zk,Abelev:2008ag} were consistent with zero within error bars.  
STAR continued measurements during the beam energy scan (BES) experiments   
and obtained positive results in particular at the lower energies~\cite{STAR:2017ckg}. 
The obtained value averaged over energy is $(1.08 \pm 0.15  \pm 0.11)\%$ 
and  $(1.38 \pm 0.30 \pm  0.13)\%$ for $\Lambda $ and $\bar\Lambda$ respectively.

With much higher statistics, the STAR collaboration has not only repeated measurements at 200GeV~\cite{Adam:2018ivw} 
but also started a systematic study on fine structures of GPE such as dependences on the centrality, the transverse momentum, 
the rapidity and also for different hyperons~\cite{Adam:2018ivw,STAR:2020xbm}, also measurements with high statistics during BES II experiments~\cite{STAR:2021beb}. 
Besides, the ALICE Collaboration has carried out measurements at LHC energies~\cite{ALICE:2019onw}, 
the HADES Collaboration at GSI has released preliminary results in the low energy region~\cite{Kornas:2020qzi}. 
Measurements at other places such as NICA at JINR are also planned. 
The results of all measurements yet available on global hyperon polarizations confirm the energy dependence of GPE observed by STAR in Ref.~\cite{STAR:2017ckg}. 

The data~\cite{STAR:2017ckg,Adam:2018ivw,ALICE:2019onw,Kornas:2020qzi,STAR:2020xbm,STAR:2021beb} available on the global hyperon polarization 
from different experiments show consistent energy dependence and other regularities. 
The order of magnitude appears to be consistent with qualitative theoretical estimates~\cite{Gao:2020lxh} based on QCD spin orbit coupling~\cite{Liang:2004ph,Gao:2007bc}. 
Different phenomenological approaches have been proposed for quantitative descriptions~\cite{Karpenko:2016jyx,Xie:2017upb,Sun:2017xhx,Li:2017slc,Liu:2019krs,Ivanov:2020qqe} 
and the consistency to the data is quite satisfactorily. 
%For a summary, see e.g. plenary talks at recent Quark Matter conferences. 
 
\subsection{Challenges and/or opportunities}

However, there are also places where experimental results are far away from theoretical expectations in particular when the fine structures are concerned. 
They are considered as cutting edge problems and/or challenges to the theory and are places that lead to rapid developments in the field. 
In the following, we present three of such examples. 

(1) The vector meson spin alignment: 
It was expected vector meson spin alignment $\rho_{00}$ is a quadratic effect of $P_q$ thus is much smaller than that of hyperon polarization. 
Experiments have been carried out by STAR and ALICE Collaborations~\cite{ALICE:2019aid,Singha:2020qns} for $K^*$ and $\phi$ mesons. 
The results show that they are indeed smaller than $1/3$. 
However, their magnitudes are much larger than that expected from $P_H$. 
Furthermore, there are also significant differences between $\rho_{00}^K$ and $\rho_{00}^\phi$, and between those at RHIC and LHC energies. 
These features are rather surprising and have attracted much theoretical efforts currently~\cite{Sheng:2019kmk,Sheng:2020ghv,Xia:2020tyd}. 

(2) The local polarization effect: 
After deriving the relationship between the polarization and the vorticity from the quantum kinetic theory, it was realized that~\cite{Gao:2012ix}
local polarization effect may exist duo to local vorticity in QGP. 
Soon after, Refs.~\cite{Voloshin:2017kqp,Becattini:2017gcx} reached the same conclusion and proposed that this could tested by measuring the azimuthal angle dependence of 
the longitudinal hyperon polarization in HIC in the transverse plane. 
Theoretical predictions have been made by studying the local vorticity in hydrodynamic model and also using Monte-Carlo event generator.
The STAR Collaboration has carried out the measurements~\cite{STAR:2019erd} but the azimuthal angle dependence is very different from the theoretical expectations. 
The signs in given quadrate are different between theory and experiment. 
This is sometimes called the sign problem and raised much studies currently~\cite{Wu:2019eyi,Fu:2021pok,Becattini:2021suc,Becattini:2021iol}. 

(3) GPE in the low energy limit: 
In this energy, QCD phase transition is expected and the GPE could be considered as one of the signatures. 
The HADES preliminary results~\cite{Kornas:2020qzi} at such low energy seem to suggest that GPE continue to rise with decreasing energy. 
This is a place worthwhile for further studies and theoretical efforts have already been made and/or are underway~\cite{Ivanov:2017dff,Kolomeitsev:2018svb,Deng:2020pep,Deng:2020ygd,Deng:2021miw}.

\section{Summary and Outlook}

To summarize, there is a great advantage to study spin physics in HIC because the reaction plane can be determined experimentally.
High energy HIC is usually non-central thus the colliding system and the produced partonic system 
QGP carries a huge global OAM as large as $10^5\hbar$ in Au-Au collisions at RHIC energies.
Due to the spin-orbit coupling in QCD, such huge OAM can be transferred 
to quarks and anti-quarks thus leads to a globally polarized QGP.  
The global polarization of quarks and anti-quarks manifest itself as the global polarization of 
hadrons such as hyperons and vector mesons produced in HIC. 

The theoretical prediction~\cite{Liang:2004ph} 
and discovery by the STAR Collaboration~\cite{STAR:2017ckg}  
open a new avenue to study properties of QGP and a new direction in high energy heavy ion physics. 
Similar measurements have been carried in other experiments such as those 
by ALICE Collaboration at LHC~\cite{ALICE:2019onw} and HADES at GSI~\cite{Kornas:2020qzi}. 
The STAR BES II will provide an excellent opportunity to study GPE in HIC and 
we expect new results with higher accuracies in next years.  
There is also plan in NICA experiment at JINR. 

The experimental efforts in turns further inspire theoretical studies. 
The rapid progresses and continuous studies along this line lead to a very active research direction 
-- the Spin Physics in HIC in the field of high energy nuclear physics. 
Among the most active aspects, we have in particular:  
(i) {GPE phenomenology} including experimental and different phenomenological approaches; 
(ii) {Spin-vortical effects in strong interacting system}; 
(iii) {Spin-magnetic effects in HIC} including 
(a) the fine structure of GPE of different hadrons,  
(b) chiral magnetic effect and 
(c) spin-electromagnetic effects in ultra-peripheral collisions (UPC) in HIC; 
and 
(iv) {Spin transport theory in relativistic quantum system} 
to derive GPE, describe the spin transport, calculate the polarization and other related spin effects directly from QCD. 

We apologize for not being able to cover all the different aspects in this direction. 
We thank in particular many collaborators for excellent collaborations on this subject. 
This work was supported in part by 
the National Natural Science Foundation of China (Nos. 11890713 and 11890700).

%\begin{thebibliography}{9}
%\bibitem{cp} The abbreviation for JPS Conference Proceedings should be ``JPS Conf. Proc." in the reference list.
%\bibitem{jpsj} The abbreviation for the Journal of the Physical Society of Japan should be ``J. Phys. Soc. Jpn." in the reference list.
%\bibitem{ptep} The abbreviation for the Progress of Theoretical and Experimental Physics should be ``Prog. Theor. Exp. Phys." in the reference list.
%\bibitem{instructions} More abbreviations of journal titles are listed in ``Instructions for Preparation of Manuscript", which is available at our Web site (http://jpsj.jps.or.jp).
%\bibitem{format} F. Author, S. Author, and T. Author, Abbreviated journal title \textbf{volume in bold face}, initial page or article number (year of publication).
%\end{thebibliography}

%%%%%%%%%%%%%%%%%%%%%%%% referenc.tex %%%%%%%%%%%%%%%%%%%%%%%%%%%%%%
% sample references
% %
% Use this file as a template for your own input.
%
%%%%%%%%%%%%%%%%%%%%%%%% Springer-Verlag %%%%%%%%%%%%%%%%%%%%%%%%%%
%
% BibTeX users please use
% \bibliographystyle{}
% \bibliography{}
%

\end{document}